%Paper: hep-th/9410052
%From: WITTEN@sns.ias.edu
%Date: 07 Oct 1994 14:35:20 -0400 (EDT)

\input harvmac
\newcount\figno
\figno=0
\def\fig#1#2#3{
\par\begingroup\parindent=0pt\leftskip=1cm\rightskip=1cm\parindent=0pt
\baselineskip=11pt
\global\advance\figno by 1
\midinsert
\epsfxsize=#3
\centerline{\epsfbox{#2}}
\vskip 12pt
{\bf Fig. \the\figno:} #1\par
\endinsert\endgroup\par
}
\def\figlabel#1{\xdef#1{\the\figno}}
\def\encadremath#1{\vbox{\hrule\hbox{\vrule\kern8pt\vbox{\kern8pt
\hbox{$\displaystyle #1$}\kern8pt}
\kern8pt\vrule}\hrule}}

\overfullrule=0pt

%macros
%

\def\bar{\overline}

\font\zfont = cmss10 %scaled \magstep1

\def\bigone{\hbox{1\kern -.23em {\rm l}}}
\def\ZZ{\hbox{\zfont Z\kern-.4emZ}}

\Title{hep-th/9410052, IASSNS-HEP-94-75}
{\vbox{\centerline{Sigma Models }
\bigskip
\centerline{And The ADHM Construction Of Instantons}}}
\smallskip
\centerline{Edward Witten}
\smallskip
\centerline{\it School of Natural Sciences, Institute for Advanced Study}
\centerline{\it Olden Lane, Princeton, NJ 08540, USA}\bigskip
\baselineskip 18pt

\medskip

\noindent
%write abstract here

This paper is devoted to the construction of
a family of linear sigma models with $(0,4)$ supersymmetry
which should flow in the infrared to the stringy version of Yang-Mills
instantons on ${\bf R}^4$.  The family depends on the full set of
expected parameters and is obtained by using the data that appear
in the ADHM construction of instantons.
\Date{October, 1994}
%text of paper
\newsec{Introduction}

There has long been interest in finding solutions
of string theory which -- in the large scale limit -- go over
to the familiar instanton solutions of Yang-Mills field theory.
An argument using supersymmetry shows that any field theoretic
instanton
solution can be systematically corrected, by adding terms of higher and
higher order in $\alpha'$, to get a string theoretic solution \ref\who{C. G.
Callan, Jr., J. A. Harvey, and A. Strominger, ``World-brane Actions
For String Solitons,'' Nucl. Phys. {\bf B367} (1991) 60, and ``Supersymmetric
String Solitons,'' in the Proceedings, String Theory and Quantum Gravity
'91 (Trieste, 1991).}.

This perturbative approach can be made particularly explicit for those
solutions that have a simple description in field theory.
For instance, if the gauge group is $SU(2)$, then an instanton of topological
charge $k$ depends on $8k-3$ parameters (or $8k$ if one includes
the global $SU(2)$ rotations).  A sub-family depending
on $5k+4$ parameters can be described by a particularly nice ansatz
introduced by 't Hooft, while to describe the full $8k-3$ parameter
family requires the more sophisticated ADHM construction
\ref\adhm{M. F. Atiyah, V. G. Drinfeld, N. J. Hitchin, and Y. I. Manin,
``Construction Of Instantons,'' Phys. Lett. {\bf 65A} (1978) 185.}.
(The ADHM construction was originally obtained using twistor space and
algebraic geometry, but can be derived and explained in terms of differential
geometry of ${\bf R}^4$ \ref\corrigan{E. Corrigan and P. Goddard,
``Construction Of Instanton And Monopole Solutions And Reciprocity,''
Annals of Physics {\bf 154} (1984) 253.}.)
\nref\strom{A. Strominger, ``Heterotic Solitons,'' Nuclear Physics
{\bf B343} (1990) 167.}
\nref\harv{C. G. Callan, Jr., J. A. Harvey, and A. Strominger,
``Worldsheet Approach To Heterotic Instantons and Solitons,''
Nucl. Phys. {\bf B359} (1991) 611.}
At least to first order in $\alpha'$, the 't Hooft ansatz fits nicely
with string theory -- it has an elegant extension
\refs{\strom,\harv} that obeys the low
energy equations derived from the heterotic string.
  On the other hand, the special properties
of the ADHM construction  have not yet been exploited in any stringy way.
That will be done in the present paper.

\nref\martinec{E. Martinec, ``Criticality, Catastrophes, And
Compactifications,'' in {\it Physics And Mathematics Of Strings},
ed. L. Brink et. al. (World Scientific, 1990) p. 389.}
\nref\vafa{C. Vafa and N. P. Warner, ``Catastrophes And The
Classification Of Conformal Theories,'' Phys. Lett. {\bf B218} (1989) 51.}
The solutions of string theory corresponding to instantons should
be conformal field theories with $(0,4)$ supersymmetry, since in the
field theory limit (that is the limit of a large scale instanton
or equivalently $\alpha'\to 0$),
self-duality of the gauge field is the condition
for $(0,4)$ supersymmetry.  Conformal
field theories often arise as the infrared limit of theories that
are not conformally invariant; a simple linear field theory can flow
in the infrared to a very subtle nonlinear model.  For instance,
this point of view has been applied to Calabi-Yau models
\refs{\martinec,\vafa}.  To study instantons in this way, we should
consider linear sigma models with $(0,4)$ supersymmetry.
The mass terms and potentials in such a model
violate conformal invariance at the classical level;
\nref\hull{C. M. Hull, G. Papadopoulos, and P. K. Townsend,
``Potentials For $(p,0)$ and $(1,1)$ Supersymmetric Sigma Models
With Torsion,'' Phys. Lett. {\bf B316} (1993) 291.}
$(0,4)$ models with such interactions
have been studied before \hull, but we will here need a slightly new
twist.

The conclusion we get can be stated as follows: in the right context,
with the right multiplets, the condition that a $(0,1)$ model
should have $(0,4)$ supersymmetry is that the Yukawa couplings
should obey the ADHM equations!
Thus, for every instanton -- that is, every solution of the ADHM equations --
we get a $(0,4)$ sigma model that should flow in the infrared to the
solution of string theory corresponding to the given instanton.

\nref\hg{J. Harvey and J. Gauntlett, ``$S$-Duality And The Spectrum Of
Magnetic Monopoles In Heterotic String Theory,'' hep-th 9407111, EFI-94-36.}
\nref\vw{C. Vafa and E. Witten, ``A Strong Coupling Test Of $S$-Duality,''
hep-th 9408074, HUTP-94-A017, IASSNS-HEP-94-54.}
This gives a uniform treatment of all instantons in string theory, and
I believe that, at least to physicists, it will put the ADHM construction
on more familiar grounds.\foot{However, while the relation to $(0,4)$
supersymmetry makes it obvious that the gauge fields coming from the ADHM
construction are self-dual, it does not necessarily shed light on the harder
part of the ADHM construction, which is to show that all instantons
arise from this construction.  That question is simply reinterpreted
as the statement that all instantons arise from linear sigma models.}
Moreover, it may enable one to get some information about the behavior
in string theory as an instanton shrinks to zero size; we comment
on this issue in section 3.  This may be helpful in understanding
issues of $H$-monopoles and $S$-duality \refs{\hg,\vw}.
Also the treatment of instantons via linear
sigma models probably  carries over to other manifolds, such as ALE
spaces, on which there is a version of the ADHM construction
\ref\kron{P. Kronheimer, ``The Construction Of ALE Spaces As Hyper-Kahler
Quotients,'' J. Diff. Geom. {\bf 29} (1989) 665.}.

\newsec{Construction Of The Model}

\subsec{Supersymmetries And Multiplets}

We work in two dimensional Minkowski space with coordinates $\tau,\sigma$
and light cone variables $\sigma^\pm=(\tau\pm\sigma)/
\sqrt 2$; the world-sheet
metric is $ds^2=d\tau^2-d\sigma^2$.
We are interested in models with $(0,4)$ supersymmetry.
This means that there are four real right-moving supercharges.
On the space of four real supercharges one could assume a symmetry
group $SO(4)\cong SU(2)\times SU(2)$.
However, the usual $N=4$ superconformal
algebra, to which the theories we formulate should flow in the infrared,
contains only a single $SU(2)$ factor.  We will call the two $SU(2)$'s
$F$ and $F'$.  Much of the formalism is invariant under $F\times F'$,
but the eventual Lagrangians will be only $F'$-invariant.

We write the four supersymmetries as $Q^{AA'}$, $A,A'=1,2$.
Indices $A,B,C=1,2$ will
transform as a doublet under $F$, and indices
$A',B',C'=1,2$ will transform
as a doublet under $F'$.
The supercharges are real in the sense that
\eqn\realsense{Q^{AA'}=\epsilon^{AB}\epsilon^{A'B'}
Q^\dagger_{BB'},}
where $\epsilon_{AB}$ and $\epsilon_{A'B'}$ are the antisymmetric
tensors of the two $SU(2)$'s.
The supersymmetry algebra is to be
\eqn\ealsense{\{Q^{AA'},Q^{BB'}\}=\epsilon^{AB}\epsilon^{A'B'}
P^+,}
where $P^+=P_-=-i\partial/\partial \sigma^-$
is the generator of translations of $\sigma^-$.

Consider a multiplet containing bosons and fermions related by the $Q$'s.
The multiplet usually considered in conformal field theory
(assuming that the eventual Lagrangian
is to be $F'$-invariant and not $F$ invariant) is one in which the
bosons transform as $(1/2,0)$ under $F\times F'$.\foot{The reason for
this is that, in a conformal $(0,4)$ theory,
if $F'$ is the only global symmetry, as is usually the case,
it must be carried only by right-moving modes, but the
bosons have also a left-moving part.  In the models we consider below,
the bosons that are not $F'$-invariant are massive and irrelevant
in the infrared.}
We will, however, also need a multiplet in which the bosons
transform as $(0,1/2)$.  We will
refer to the two types of multiplets as standard and twisted multiplets,
respectively.
Other multiplets have been found recently.\ref\gates{J. Gates and L. Rana,
``On Extended Supersymmetric Quantum Mechanics,'' Univ. of Maryland
preprint, to appear.}

First, we write down the structure of the standard multiplet.  (We will work
in components, as I do not know how to describe these models in superspace.)
There is a bose field $X^{A Y}$, $A,Y=1,2$,
with a reality condition analogous to \realsense:
\eqn\realcon{X^{A Y}=\epsilon^{AB}\epsilon^{YZ}\overline
X_{BZ}.}
This multiplet admits the action of yet another $SU(2)$ group $H$
(which will generally not be a symmetry of the theories that we eventually
consider). $F$ acts on indices $A,B,C$ and $H$ on the indices $Y,Z$;
the $X$'s thus transform as $(1/2,0,1/2)$ under $F\times F'\times H$.
$X$ is related by supersymmetry to a right-moving \foot{$\psi_-$
is right-moving in the sense that the equation of motion
of the free $\psi_-$ field would be $\partial_+\psi_-=0$, so in the free
theory $\psi_-$ is right-moving.  A similar statement holds
for what we will later call left-moving fields.} fermi field $\psi_-^{A'Y}$
with a reality condition of the same
form as \realcon.
The supersymmetry transformation laws are
\eqn\ungo{\eqalign{\delta X^{A Y} & = i\epsilon_{A'B'}
       \eta_+^{AA'}\psi_-^{B'Y}   \cr
            \delta \psi^{A'Y} & = \epsilon_{AB}
             \eta_+^{AA'}\partial_-X^{BY},\cr}}
where $\eta_+^{AA'}$ are infinitesimal parameters.
It is easy to check the commutation relations
\eqn\tungo{[\delta_{\eta'},\delta_\eta]=-i\epsilon_{AB}\epsilon_{A'B'}
\eta_+^{AA'}\eta'{}_+^{BB'}\partial_-.}
If one wishes to consider $k$ such multiplets, one simply extends
the $Y$ index to run from $1\dots 2k$; the group $H$ becomes the symplectic
group $Sp(k)$
instead of $SU(2)\cong Sp(1)$, and the tensor $\epsilon^{YZ}$
in \realcon\ should be understood as the invariant antisymmetric
tensor of $Sp(k)$.

The structure of the twisted multiplet is the same, except that
the roles of $F$ and $F'$ are reversed.  Thus, there
is a bose field $\phi^{A'Y'}$, $A',Y'=1,2$ with a reality condition like that
in \realcon:
\eqn\bealcon{\phi^{A'Y'}=\epsilon^{A'B'}\epsilon^{Y'Z'}\overline \phi_{B'Z'}.}
The $A'$ index transforms as a doublet of $F'$, and $Y'$ is acted on by
a new $SU(2)$ group $H'$.  In addition, there is a right-moving
fermi multiplet $\chi_-
^{A Y'}$ (an $F'$ singlet, as the notation indicates),
obeying a reality condition like that in \bealcon.  The transformation laws
are obtained from \ungo\ with obvious substitutions:
\eqn\uungo{\eqalign{\delta \phi^{A'Y'} & = i\epsilon_{AB}
       \eta_+^{AA'}\chi_-^{BY'}   \cr
            \delta \chi_-^{AY'} & = \epsilon_{A'B'}
             \eta_+^{AA'}\partial_-\phi ^{B'Y'}.\cr}}
The commutation relations of \tungo\ are of course obeyed.
For $k'$ such multiplets, one  must take $H'$ to be $Sp(k')$,
take $Y'$ to run from $1\dots 2k'$ and
interpret the tensor $\epsilon^{Y'Z'}$ in the reality condition as the
invariant antisymmetric tensor of $Sp(k')$.

Of course, until we start writing Lagrangians, there is no essential
difference between $F$ and $F'$, and the two multiplets
are on the same footing.  They become distinguished once we construct
Lagrangians that are invariant under $F'$ and not $F$.

\subsec{Left-Moving Fermions}

Now we introduce left-moving fermions $\lambda_+^a$, $a=1\dots n$.
The part of the Lagrangian containing $\lambda_+^a$ is of the general
form
\eqn\partof{\int d^2\sigma\left({i\over 2}\lambda_+^a\partial_-\lambda_+^a
      -{i\over 2}\lambda_+^aG_{a\theta}\rho_-^\theta\right),}
where the components of $\rho_-$ include all
the right-moving fermions $\psi,\chi$,
and $G_{a\theta}$ is some unknown function of $X$ and $\phi$
incorporating masses and
Yukawa couplings.  The equations of motion of the
$\lambda$'s are thus
\eqn\artof{\partial_-\lambda_+^a=G^a_\theta\rho_-^\theta.}

Since I do not know how to treat the $\lambda$'s as a $(0,4)$ superfield,
we will have to study the supersymmetry transformations on-shell and by hand.
The most general supersymmetry transformation allowed for the $\lambda$'s
by dimensional analysis is
\eqn\gartof{\delta\lambda_+^a= \eta_+^{AA'}C_{AA'}^a,}
with $C$ a function of $X,\phi$.
Calculating the second variation, we get
\eqn\uartof{\delta_{\eta'}\delta_\eta \lambda_+^a =i\eta_+^{AA'}
\left({\partial C_{AA'}^a\over\partial X^{BY}}\epsilon_{B'C'}
\eta'{}_+^{BB'}
\psi_{-}^{C'Y}+{\partial C_{AA'}^a
\over\partial\phi^{B'Y'}}
\epsilon_{BC}\eta'{}_+^{BB'}\chi^{CY'}\right).}
We want to compare this to the expected result
\eqn\yartof{\eqalign{\left(\delta_{\eta'}\delta_\eta-\delta_\eta
\delta_{\eta'}\right)\lambda_+^a&=-i\epsilon_{A'B'}\epsilon_{AB}
\eta_+^{AA'}\eta'{}_+^{ BB'}\partial_-\lambda_+^a\cr &
=-i\epsilon_{AB}\epsilon_{A'B'}
\eta_+^{AA'}\eta'{}_+^{ BB'}G^a_\theta\rho^\theta.\cr}}

The condition for \uartof\ to be of the form \yartof\ -- for a suitable
$G$ -- is simply that
\eqn\piftof{\eqalign{0= & {\partial C_{AA'}^a\over\partial X^{B Y}}
+{\partial C_{BA'}^a\over\partial X^{A Y}}
 =  {\partial C_{AA'}^a\over\partial \phi^{B'Y'}}
+{\partial C_{AB'}^a\over\partial \phi^{A'Y'}}.\cr}}
If this is so, then \uartof\ and \yartof\ agree with
\eqn\ippo{G_\theta^a\rho^\theta={1\over 2}\left(\epsilon^{BD}{\partial
C^a_{BB'}\over\partial X^{DY}}\psi_-^{B'Y}+\epsilon^{B'D'}{\partial C^a_{BB'}
\over\partial \phi^{D'Y'}}\chi^{BY'}\right).}

In \gartof\ and \piftof, we have obtained a considerable amount of
information about the supersymmetry transformation law of $\lambda$.
But is there actually a Lagrangian that is invariant under this symmetry?
To answer this question, it is useful to compare to $(0,1)$ supersymmetry,
where there is a superspace formulation.  In that case, $\lambda$
is part of a multiplet $\Lambda^a=\lambda^a+\theta F^a$, where $F$ is an
auxiliary field that eventually (by equations of motion) becomes a function
of the other bosons.  The supersymmetry transformation law of $\lambda$
is
\eqn\miccon{\delta\lambda^a=\eta F^a}
and the potential energy of the theory is
\eqn\iccon{V={1\over 2}\sum_aF^aF^a.}

Comparing \miccon\
to \gartof, it is clear that $F$ corresponds to a component of $C$.
If we pick any real $c$-number $c^{AA'}$, normalized so
\eqn\uj{\epsilon_{AB}\epsilon_{A'B'}c^{AA'}c^{BB'}=1,}
then we can make a $(0,1)$ supersymmetric model in which the transformation
laws of $\lambda$ are the ones given above specialized to $\eta^{AA'}
=\eta c^{AA'}$, $\eta$ being an anticommuting parameter of
$(0,1)$ supersymmetry.  The potential of this theory will be
\eqn\nuj{V={1\over 2}\sum_a C^aC^a}
where
\eqn\ono{C^a=c^{AA'}C_{AA'}^a.}

To get $(0,1)$ supersymmetry with the transformation laws in \gartof,
we need the potential in \nuj\ to be independent of $c$, ensuring
that the {\it same} Lagrangian is invariant under all four supersymmetries.
The condition for this is that
\eqn\pbon{0=\sum_a\left(C_{AA'}^aC_{BB'}^a+C_{BA'}^aC_{AB'}^a
\right).}

We now have enough information to determine what the $(0,4)$
Lagrangian is, if there is one.  The
Yukawa couplings are given in \partof\ and the potential in \nuj.
The Lagrangian is therefore
\eqn\impo{L=L_{kin}-{i\over 4}\int d^2\sigma \lambda_+^a
\left(\epsilon^{BD}{\partial
C^a_{BB'}\over\partial X^{DY}}\psi_-^{B'Y}+\epsilon^{B'D'}{\partial C^a_{BB'}
\over\partial \phi^{D'Y'}}\chi_-^{BY'}\right)-{1\over 8}\int d^2\sigma
\epsilon^{AB}\epsilon^{A'B'}C^a_{AA'}C^a_{BB'},}
where $L_{kin}$ is the free kinetic energy for all fields.
$(0,4)$ supersymmetry does hold when
\piftof\ and \pbon\ are valid since \pbon\ insures that the Lagrangian
is invariant under all the supersymmetries and \piftof\ ensures that
these generate the global $(0,4)$ algebra.

\subsec{The ADHM Equations}

Now let us make the meaning of \piftof\ and \pbon\ more explicit.
\piftof\ can be solved quite explicitly; the general solution (raising
and lowering indices with the $\epsilon$ symbols) is
\eqn\pimo{C_{AA'}^a=M_{AA'}^a+X_{AY}N^{a\,Y}_{A'}
 +\phi_{A'}{}^{Y'}D_{AY'}^a+X_{A}{}^{Y}\phi_{A'}{}^{Y'}E^a_{YY'}  }
where $M,N,D$, and $E$ are independent of $X$ and $\phi$.  Since $C$
thus contains only terms at most bilinear in $X$ and $\phi$, the potential
$V\sim C^2$ at most contains terms of degree $(2,2)$.

It is straightforward now to impose the equations \pbon, giving
a finite system of equations for the finite set of coefficients in
$M,N,D,E$.  However, to imitate as much as possible the structure
of superconformal $(0,4)$ theories, we want to impose invariance
under $F'$ (or $F$).  $F'$ invariance simply means that $M=N=0$,
since $M_{AA'}^a$ and $N_{A'Y}^a$ transform as doublets of $F'$
(which acts on the $A'$ subscript).  Thus in the $F'$ invariant
case, we can succinctly write
\eqn\timo{C_{AA'}^a=B_{AY'}^a\phi_{A'}{}^{Y'},}
where $B$ is linear in $X$ (and independent of $\phi$).
The equation \pbon\ can then be conveniently written
\eqn\mimo{\sum_a\left(B_{AY'}^aB_{BZ'}^a+
B_{BY'}^aB_{AZ'}^a\right)=0.  }
Thus, the $F'$-invariant $(0,4)$ theories are simply determined by
a tensor $B$, linear in $X$, and obeying \mimo.  Of course, it may
also be interesting to study the theories that arise if one does not
assume either $F$ or $F'$ invariance.

Now let us discuss the physical significance of this model.
Since $C$ is
homogeneous and linear in $\phi$, the potential $V$ is
homogeneous and quadratic in $\phi$; in particular, it vanishes at
$\phi=0$, for any $X$.  Therefore, the $X$'s are massless fields,
but the $\phi$'s are massive; indeed, being quadratic in $\phi$,
$V$ can be interpreted as an $X$-dependent mass term for the $\phi$'s.
If the number $n$ of components of $\lambda$ is big enough and
the solution $B$ of \mimo\ is sufficiently generic, then for every
value of $X$, all of the $\phi$'s are massive.  (From what we will presently
say, that this generically happens
follows from standard theorems about the ADHM construction.)
So the massless
particles are precisely the $X$'s, and the space of vacua is ${\cal M}=
{\bf R}^{4k}$,
parametrized by the $X$'s.  (We recall that $k$ and $k'$ are the
numbers of $X$ and $\phi$ multiplets.)  At the classical level, the
metric on ${\cal M}$ is read off from the classical Lagrangian after
setting the $\phi$'s to zero, and (assuming that we started with the free
kinetic energy for all fields), it is just the flat metric on ${\bf R}^{4k}$.
Of course, the $(0,4)$ supersymmetry endows ${\cal M}$ with a hyperkahler
structure.

What about the fermions?  On ${\cal M}$, that is, at $\phi=0$, the
structure of the Yukawa couplings is particularly simple; it reduces
to
\eqn\jumper{ \sum_a \lambda_+^aB^a_{AY'}\chi_-^{AY'}.}
Thus, the fermionic partners $\psi_-$ of $X$ are all massless,
as one would expect from $(0,4)$ supersymmetry.  If the number $n$
of components of $\lambda_+^a$ is bigger than the number $4k'$ of
components of $\chi_-^{AY'}$, then generically all components of $\chi_-$
get mass (in fact, by supersymmetry this is true precisely when all
components of $\phi$ are massive), and $N=n-4k'$ components of $\lambda_+$
are massless.

Let $v_i{}^a$, $i=1\dots N$, be a basis of the massless components
of $\lambda_+$, that is the solutions of $\sum_av_i^aB^a_{AY'}=0$.
Choose the $v$'s to be orthonormal, that is $\sum_a
v_i{}^av_j{}^a=\delta_{ij}$.
Of course, the $v_i^a$ are $X$-dependent, as the tensor $B$ is $X$-dependent
(in fact, linear in $X$).  Orthonormality determines $v_i^a$ up to an
$X$-dependent $SO(N)$ transformation on the $i$ index; this will be
interpreted presently as the gauge invariance of the low energy theory.
One can set to zero the massive modes in $\lambda_+$ by writing
\eqn\hunny{\lambda_{+}^a=\sum_{i=1}^N v_i^a\lambda_{+\,i},}
where $\lambda_{+\,i}$ are the massless left-moving fermions.

To write, at the classical level, the effective action for the massless
modes, one sets to zero the massive fields $\phi,\chi$, and makes
the ansatz \hunny\ for $\lambda_+$.  In particular, the kinetic energy
for $\lambda_+$ becomes
\eqn\inny{{i\over 2}
\int d^2\sigma \,\sum_{i,j,a}\left(v_i^a\lambda_{+\,i}\right)
\partial_- \left(v_j^a\lambda_{+\,j}\right)
         = {i\over 2}\int d^2\sigma\,\sum_{i,j}
\left\{\lambda_{+\,i}\left(\delta_{ij}\partial_-
                         +\partial_-X^{BB'}A_{BB'\,ij}\right)\lambda_{+\,j}
\right\},}
with
\eqn\jinny{A_{BB'\,ij}= \sum_av_i^a{\partial v_j^a\over\partial X^{BB'}}.}
\inny\ is the standard sigma model expression for couplings of left-moving
fermions to space-time gauge fields, the gauge field being given in \jinny.

$(0,4)$ supersymmetry means that the gauge field in \jinny\ must be
compatible with the hyperkahler structure of ${\bf R}^{4k}$
(that is, its curvature is of type $(1,1)$ for each of the complex
structures).  For $k=1$, this reduces to a more familiar statement:
the gauge field is an instanton; its curvature is anti-self-dual.

In fact, for $k=1$, what we have just obtained
is simply the ADHM construction of instantons.
According to the ADHM construction, an instanton of $SO(N)$, with
instanton number $k'$, is given by a tensor $B$, linear in $X$,
that obeys \mimo, and has a further non-degeneracy condition which
simply asserts that the components of $\phi$ are all massive for all $X$.
Moreover, \jinny\ is the standard ADHM formula for the gauge field, in terms
of $B$.  The definition of the gauge field is actually perhaps better
explained without formulas; the bundle $E$ of massless fermions is a subbundle
of the trivial bundle $E_0$ of all fermions $\lambda_-^a$, and the gauge
connection on $E$ is the connection on $E$ induced from the trivial
connection on $E_0$.  In formulas, that corresponds to \jinny.

The ADHM theorem further asserts that two instantons
are equivalent if and only if the two $B$ tensors can be mapped
into each other by the action of $SO(n)\times Sp(k')$ (which comes
from the linear action of this group
on the $a$ and $Y'$ indices of $\lambda_+{}^a$ and $\phi^{A'Y'}$).

\nref\phases{E. Witten, ``Phases Of $N=2$ Models In Two Dimensions,''
Nucl. Phys. {\bf B403} (1993) 159.}
The ADHM construction gives rise to a partial compactification
of instanton moduli space in which one drops the non-degeneracy
condition and permits instantons for which $\phi$ is massless at some
values of $X$; this corresponds to allowing some instantons to shrink
to zero size.  One wonders if this compactification is relevant
to string theory, especially since linear sigma models have a good
record of predicting
correctly the moduli spaces of conformal field theories including
the subtle behavior associated with singularities (for instance,
see \phases).  The partial compactification of moduli space that comes
from the ADHM construction is actually the analog for ${\bf R}^4$ of
the compactification that has been (rather optimistically)
used in testing $S$-duality of $N=4$ super Yang-Mills theory for strong
coupling \vw.

\newsec{Instanton Number One}

In this section, I will make this construction
somewhat more explicit by describing
precisely the instanton number one solution in this language.
Of course, there is no essential novelty here; the formulas are well
known in the ADHM literature.  Writing them out here
may nevertheless be useful.

The basic one instanton solution arises for gauge group $SU(2)$,
but of course can be embedded in any larger gauge group.  In the formulation
above, the natural gauge group is $SO(N)$.  We will take $N=4$
and use the embedding of $SU(2)$ in $SO(4)$
given by the decomposition $SO(4)\cong SU(2)\times SU(2)$.  Thus,
we identify the gauge group of the one instanton solution with one of the
$SU(2)$'s in $SO(4)$; the second $SU(2)$ -- call it $K$ -- is a (rather
trivial) global symmetry group of the solution.

Now, let us consider the full symmetry group of the sigma model corresponding
to this solution.  Any $(0,4)$ sigma model of the type we are considering
has a global symmetry group $F'\cong SU(2)$.  In addition, the one instanton
solution happens to be invariant under rotations \foot{In field
theory, the instanton solution is actually invariant in addition
under certain conformal
transformations of ${\bf R}^4$, but this depends on the conformal
invariance of the self-dual Yang-Mills equations and is lost in string
theory.}  of ${\bf R}^4$.
The rotation group is $SO(4)\cong SU(2)_L\times SU(2)_R$.
So altogether, the global symmetry of the $(0,4)$ sigma model that we are
looking for is $F'\times K\times SU(2)_L\times SU(2)_R\cong SU(2)^4$.

What are the fields supposed to be?  We want $k=1$ to get ${\bf R}^4$
as space-time, and $k'=1$ so that the instanton number will be one.
There are therefore four $X$'s, $X^{AY}$, and four $\phi$'s, $\phi^{A'Y'}$.
On the $X$'s and $\phi$'s there is a natural action, noted in the last
section, of $F\times F'\times H
\times H'\cong SU(2)^4$.
It is tempting to try to identify $F\times F'\times H\times H'$ with
the isomorphic group $F'\times K\times SU(2)_L\times SU(2)_R$, and this
indeed proves to be correct.  Thus, while the general instanton-related
$(0,4)$ model of the previous section has only $F'$ symmetry, the
one instanton solution (embedded as we have described in $SO(4)$)
proves to have the full $F\times F'\times H\times H'$ symmetry.

To achieve this symmetry in the Lagrangian, one needs an appropriate
action of $F\times F'\times H\times H'$ on the left-moving fermions
$\lambda_+$.  The number $n$ of $\lambda_+$ components is 8
(so that $N=n-4k'=4$), and with a little experimentation
(or by comparing to the standard ADHM description of the basic
instanton) one finds that the $\lambda_+$'s must be taken to transform as
$(1/2,0,0,1/2)\oplus (0,0,1/2,1/2)$.  The $\lambda_+$ fields are thus
naturally written as $\lambda_+^{AY'}$, $A,Y'=1,2$ and $\lambda_+^{YY'}$, $Y,Y'
=1,2$, with the usual sort of reality condition
\eqn\turtag{\eqalign{\lambda_+^{AY'}&=\epsilon^{AB}\epsilon^{Y'Z'}\bar\lambda
_{+\,BZ'}\cr
 \lambda_+^{YY'}&=\epsilon^{YZ}\epsilon^{Y'Z'}\bar\lambda_{+\,ZZ'}.\cr}}
Note that the total number of components of $\lambda_+$ is indeed
$2\times 2+2\times 2=8$.

For this choice of $\lambda_+$'s, the coupling tensor $C^a_{BB'}$
has two pieces, $C^{AY'}{}_{BB'}$ and $C^{YY'}{}_{BB'}$.  For these
we make (up to inessential rescalings)
the most general ansatz compatible with \pimo\ and the symmetries:
\eqn\hurtag{\eqalign{C^{YY'}{}_{BB'} & = X_B{}^Y\phi_{B'}{}^{Y'}\cr
                     C^{AY'}{}_{BB'} & = {\rho\over \sqrt 2}
\delta^A{}_B\phi_{B'}{}^{Y'}.\cr}}
Here $\rho$ is a real number that will be interpreted as the instanton
scale parameter.
It is easy to verify the ADHM equations \mimo.
One may readily compute from \impo\ that the potential is
\eqn\urtag{V={1\over 8}(X^2+\rho^2)\phi^2}
where
\eqn\wurtag{X^2=\epsilon_{AB}\epsilon_{YZ}X^{AY}X^{BZ},~~~~
\phi^2=\epsilon_{A'B'}\epsilon_{Y'Z'}\phi^{A'Y'}\phi^{B'Z'}.}
The expected symmetry under $F\times F'\times H\times H'$, which acts
by independent rotations of $X$ and $\phi$,  is manifest
in \urtag.

To extract the standard formula for the one instanton solution
from these expressions, we simply need to write down the low energy
action for the massless fermions.  The following ansatz exhibits four
massless modes of $\lambda_+$:
\eqn\yuru{\eqalign{\lambda^{YY'} & = {\rho\zeta_+^{YY'}\over
                                     \sqrt{\rho^2 +X^2}}\cr
                   \lambda^{AT'} & =-{\sqrt 2 X^A{}_Y\zeta_+^{YY'}\over
               \sqrt{\rho^2+X^2}}.\cr}}
With this ansatz, the free kinetic energy of the $\lambda_+$'s (the first
term in \partof) can be rewritten using the following formula:
\eqn\wuru{\eqalign{
\lambda_{+\,YY'}\partial_-\lambda_+^{YY'}&+\lambda_{+\,AY'}\partial_-
\lambda_+^{AY'} \cr &
=\zeta_{+\,YY'}\partial_-\zeta_+^{YY'}
-\zeta_{YY'}{{1\over 2}\epsilon_{AB}\left(X^{AY}\partial_-X^{BZ}
                         +X^{AZ}\partial_-X^{BY}\right)\over
 X^2+\rho^2}\zeta_{+\,Z}{}^{Y'}.\cr}}
In the last term one sees the standard instanton gauge field.

\nref\howe{P. S. Howe and G. Papadopoulos, ``Finiteness of $(4,q)$ Models
With Torsion,'' Nucl. Phys. {\bf B289} (1987) 264, ``Further Remarks
On The Geometry Of Two-Dimensional Nonlinear $\sigma$ Models,'' Class.
Quantum Grav. {\bf 5} (1988) 1647.}
This completes what I will say about the classical sigma model describing
the instanton.  To obtain the stringy corrections to the solution, one
must consider the renormalization group flow to a (presumed) $(0,4)$
conformal field theory in the infrared.  This certainly entails
integrating out the massive fields rather than simply setting them to zero.
Whether in addition it is necessary to consider renormalization group
flow of the massless fields depends on whether the $(0,4)$ action that
one gets by integrating out the massive fields is automatically conformally
invariant.  It has been argued that massless $(0,4)$ theories
of the general type under discussion here are automatically conformally
invariant under some conditions \howe, but there are limitations on this
argument \refs{\howe,\harv}.
I will only make two observations here about the renormalization group flow:

(1) In integrating out the massive fields, the expansion
parameter is easily seen to be $1/(X^2+\rho^2)$; thus perturbation
theory is accurate for all $X$ if $\rho$ is large, and for large $X$ even
if $\rho $ is small.

(2) Despite being super-renormalizable, it appears that when formulated
on a curved two-manifold, the theory has a one-loop logarithmic divergence
proportional to the world-sheet curvature $R$,
if minimally coupled to the world-sheet metric.  It is conceivable that
to avoid this, one should add an extra coupling, perhaps of the form
$R\phi^2$, when working on a curved world-sheet.  This would spoil the
$\rho=0$ symmetry between $X$ and $\phi$ that is mentioned below.

\subsec{Behavior For $\rho=0$.}

It is interesting to ask what happens if $\rho$ is small -- in
fact, for $\rho\to 0$.  This is the regime in which the deviations of
the stringy instanton from the field theoretic instanton should be large.
In \refs{\strom,\harv},
an interesting proposal was made for the structure of the solution
at $\rho=0$: the low energy string-derived field equations were solved
for arbitrary $\rho$, and it was seen that for $\rho=0$ the solution
develops a semi-infinite tube that joins to the rest of space-time near $X=0$.
This picture is not guaranteed to be correct because it is based on
solving low energy equations that are not valid near $X=0$ when $\rho=0$.
\foot{The picture is supported by the fact, noted in \harv, that
the semi-infinite tube does indeed correspond to an exact conformal
field theory.  (This is true for $(0,4)$ as well as for the $(4,4)$ case
discussed more thoroughly in \harv.)
However, one does not know if the tube is stable.}
{}From the mean field theory developed in the present paper, one gets
a somewhat different picture of what happens at $\rho=0$. (This picture
is also not guaranteed to be correct since we do not have precise control
on the renormalization group flow.)

First of all, at $\rho=0$, the
massless fermions, in view of the above formulas,
are simply the $\lambda_+^{AY'}$; they are exactly decoupled from the other
fields, so the gauge connection is zero.  This statement agrees with the
picture of \refs{\strom,\harv} concerning what happens to the gauge
fields for $\rho=0$.
The space-time looks somewhat different, however.
The structure of the potential energy
\eqn\fifo{V={1\over 8}(X^2+\rho^2)\phi^2}
indicates that what happens at $\rho=0$ is that the theory develops
a second branch of classical vacua:
in addition to the usual branch, $\phi=0$ with any $X$,
one has a second branch, $X=0$ with any $\phi$.  Field theory is
a good approximation for large $X$ on the first branch, or for large
$\phi$ on the second branch.  Indeed, at $\rho=0$ there is a ${\bf Z}_2$
symmetry that exchanges $X$ and $\phi$ (unless this is ruined by
an $R\phi^2$ term mentioned above).  The two branches are two
asymptotically flat and empty space-times, connected by a ``worm-hole''
region
near $X=\phi=0$ where stringy effects are important.

Or alternatively,
it may be that under the renormalization group flow to the infrared,
the two branches become disconnected.  If in renormalization
group flow the two branches become infinitely
far apart, each separated from $X=\phi=0$ by a semi-infinite tube,
then it might be that the picture of
\refs{\strom,\harv} corresponds to the $X$ branch.
In any event, the meaning of the $\phi$ branch is
rather mysterious.

\listrefs
\end